\documentclass[acmtog,nonacm]{acmart}

\usepackage{booktabs} 
\usepackage{bm}
\usepackage{wrapfig}
\usepackage[
raggedright
]{sidecap}   
\sidecaptionvpos{figure}{t}

\acmJournal{TOG}
\setcopyright{none}

\newcommand{\secref}[1]{Sec.~\ref{#1}}

\definecolor{darkgreen}{rgb}{0.0, 0.7, 0.0}

\newcommand{\projectname}[0]{Neuroll}

\title{\projectname : Real-Time Neural Strand-Based Hair Simulation via Simulator-in-the-Loop Unrolling}

\author{Gene Wei-Chin Lin}
\orcid{0009-0006-5492-9941}
\email{genelin1211@gmail.com}
\affiliation{%
 \institution{Meta}
 \city{Vancouver}
 \country{Canada}}

\author{Jessica Jia-En Lee}
\orcid{0009-0008-7725-4320}
\email{nn.jessica.lee@gmail.com}

\author{Yu Ju (Edwin) Chen}
\orcid{0000-0002-7540-6246}
\email{edwinchenyj@gmail.com}
\affiliation{%
 \institution{Meta}
 \country{USA}}

\author{Egor Larionov}
\orcid{0000-0002-9900-3150}
\email{elrnv@meta.com}
\affiliation{%
\institution{Meta Reality Labs}
\country{USA}}

\author{Tuur Stuyck}
\orcid{0000-0003-1892-2137}
\email{tstuyck@gmail.com}
\affiliation{%
 \institution{NVIDIA}
 \country{USA}}

\begin{abstract}

Time integration has been the cornerstone of physics-based animation that enables the simulation of complex interactions between rigid and deformable objects, including the motion of hair. Despite recent advances with optimized time integration that enabled thousands of hair strands to be simulated in real time, achieving the same performance on commodity hardware remains infeasible due to the computational demands of resolving complex dynamics and interactions between thousands of individual strands. With the rise of learning-based techniques, the offload of time integration to neural networks helps to achieve significant performance gains, making these approaches suitable for real-time applications such as gaming and virtual avatars. However, state-of-the-art neural techniques tend to produce less physically plausible motion and oftentimes fail to generalize to out-of-distribution scenarios. Inspired by classical time integrators, we design a neural counterpart that mirrors their input-output formulation -- taking previous hair states, material stiffness, and collision geometry as the inputs for the neural time integrator, which is then trained via a self-supervised, simulator-in-the-loop method with randomized unrolling horizons. By formulating training in each strand's local coordinate frame, we obtain a network that generalizes across multiple dimensions, including hairstyle, material property, body motion, and body type. Our method inherits the benefits of a strand-based neural simulator, and hence is density-independent, lightweight, memory-efficient, and performant. Our neural hair integrator produces stable long-horizon rollouts and can be naturally extended to support quasi-static simulation simply by resetting hair states.

\end{abstract}



\begin{document}

\begin{teaserfigure}
  \includegraphics[width=\textwidth]{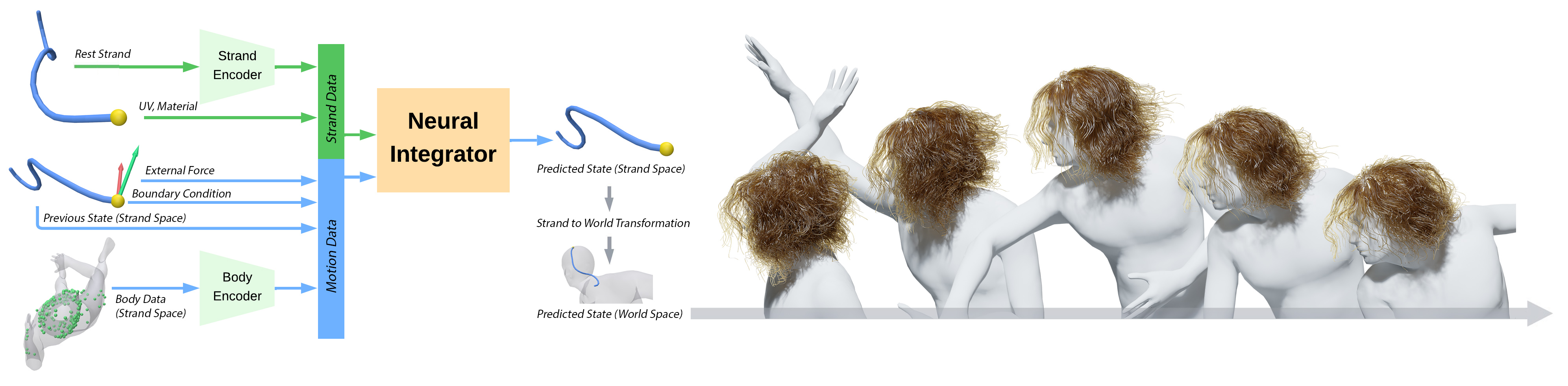}
  \caption{\projectname{}\ is a strand-based neural time integrator that predicts quasi-static and dynamic hair deformation with long horizon rollouts. Our method operates in a strand's local coordinate frame, conditions on a representation-agnostic body field, and uses simulator-in-the-loop unrolling during training, and hence generalizes across hairstyle, material property, body motion, and body type, while producing dynamic motion visually comparable to classical simulation at a fraction of the cost.
  } \label{fig:teaser}
\end{teaserfigure}

\maketitle

\section{Introduction} \label{sec:intro}

Strand-based representation of hair has become commonplace and proves to be important in numerous applications such as films, gaming and virtual reality. While strand-based groom creation is itself challenging, often requiring a mix of procedural processes and geometric manipulations from artists~\cite{lin2018digipro} or automatic reconstruction from images and videos~\cite{sklyarova2023neural, zakharov2024human}, generating realistic hair animation poses an even greater challenge. Strand-based hair motion significantly elevates visual quality and perceived authenticity, in both single-user experiences and multiple-user interactions. 

Fundamentally, time integration drives hair animation by solving PDEs at each time step. It computes the next hair state by factoring in previous states, boundary conditions, collisions, and external forces. While time integration has been well established for decades in computer graphics, animation, visual effects and gaming, with numerous methods proposed for hair simulation~\cite{bertails2011, kaufman2014adaptive, Hsu2025, daviet2023interactive}, the quality of hair motion is still bounded by the available computational resources. To fit within strict computational limits while preserving artistic intent, practitioners often resort to reduced degrees of freedom or simplified geometries, such as guide strands and mesh-based hair. However, even with significant simplification, additional measures such as substepping are still needed to maintain motion quality, further increasing computational cost.

Recent neural approaches such as GroomGen~\cite{zhou2023groomgen}, Quaffure~\cite{stuyck2024quaffure}, and Neuralocks~\cite{lin2026neuralocks} address the challenge by offloading the complex computations to neural networks and generating hair deformations in either latent or full space. GroomGen trains a network on limited synthetic data produced by artists via simulation, requiring significant manual effort and data handling, making it less generalizable across hairstyles and motion ranges. Quaffure learns a mapping from body pose to hair deformation in full space, albeit only for quasi-static cases. Neuralocks uses a strand-based network and introduces an inertia loss in the training process, enabling dynamic strand simulation and generalization while maintaining a low memory footprint and high inference efficiency. However, the inertia loss is highly sensitive to its weighting, trading overshooting against excessive stiffness with no satisfactory middle ground. Moreover, none of these approaches feeds previous hair states to the network as input; dynamics are induced only through the training objective, so the resulting motion decays over long rollouts (Table 1, $\mathcal{E}_{dyn}$).

In cloth simulation, several neural approaches avoid precomputed datasets by minimizing physical energy potentials directly during training, mapping a body motion sequence to next-step states~\cite{bertiche2022neural, santesteban2022snug}. HOOD~\cite{grigorev2023hood} leverages a graph-based garment representation that is processed by a graph neural network to predict nodal accelerations. Once trained, the network autoregressively generates dynamic garment motions over long time horizons while exhibiting good generalization. Transferring this recipe to hair is, however, not trivial due to the worse conditioned elastic energy of the Cosserat rod models used for offline hair simulation~\cite{sca2016cosserat, Hsu2025} including in Neuralocks. This has been observed to significantly degrade convergence of the network training, rendering rod-based formulations impractical for training a neural hair integrator.

Supervised, data-driven approaches such as GroomGen~\cite{zhou2023groomgen} avoid this difficulty by regressing to precomputed simulations, yielding a well-conditioned objective. In exchange, they require computationally demanding physics simulation, repeated for every hairstyle, body motion, body type, and hair material. Simulated trajectories can also differ substantially between similar body motion, making the regression target noisy~\cite{bertiche2022neural}. Moreover, due to the nature of physics simulation, quasi-static and dynamic ground truth must be generated and trained on separately, doubling the data requirements.

To avoid the overshooting–stiffness trade-off of the inertia loss and the poor training convergence caused by rod-based elastic potentials, we first represent all network inputs in a strand’s local space, then replace both terms with a positional loss between the model prediction and ground truth from a hair simulator, using unrolling~\cite{list2024differentiability} with a randomized horizon. Our method requires no data generation or curation, and uses neither GRUs nor GNNs: because the integrator consumes explicit physical states rather than pose history or a learned hidden state, its input stays compact, and inference-time behavior (stiffness, boundary conditions, and additional external forces) is controlled by manipulating those inputs directly. Once trained, the strand-based network predicts hair states at the next time step. It remains stable over long horizons and generalizes across hairstyles, body shapes, and body motions, producing substantially more dynamic motion than prior neural methods (Table 1, $\mathcal{E}_{mr}$). In addition, owing to our training strategy, the same network can be used to predict quasi-static (draped) hair deformation, eliminating the need for separate quasi-static and dynamic training. In summary, we contribute:

\begin{itemize}
    \item a training strategy for dynamic neural hair that combines simulator-in-the-loop positional supervision with randomized-horizon unrolling, requiring no precomputed dataset.
    \item a strand-local representation of network inputs, which keeps the input space small and smooth, yielding more dynamic hair motions and high generalizability.
    \item a representation-agnostic body signal for collision response, allowing the same trained network to be driven by different body representations without retraining.
    \item a single neural integrator that performs both quasi-static and dynamic simulation, selected at inference without mode-specific training.
\end{itemize}

Together, these yield, to our knowledge, the first neural strand integrator that stays dynamic over thousands of autoregressive frames (Table~\ref{tab:runtime_performance_table}, $\mathcal{E}_{dyn}$).

\section{Related Work} \label{sec:related_work}

\subsection{Physics-based Simulation} 
Physics-based hair simulation has evolved across several structural representations, ranging from early cantilever beam models \cite{anjyo1992simple, daldegan1993integrated} to volumetric formulations \cite{hadap2001modeling}, super-helices \cite{bertails2006}, Cosserat rod dynamics \cite{pai2002strands, sca2016cosserat}, discrete elastic rods \cite{bergou2008discrete}, mass-spring networks \cite{selle2008}, and explicit mesh representations \cite{yuksel2009hair}. Each representation balances physical fidelity, computational overhead, and numerical stability differently. Handling inter-strand interactions and collisions remains a central challenge, leading to specialized collision handling techniques \cite{choe2005haircontact, daviet2011, bertails2011, kaufman2014adaptive}.

To enhance visual realism and numerical accuracy, subsequent research addressed inverse simulation problems and artificial sagging artifacts \cite{derouet2013inverse, twigg2011sagfree, hsu2023sag}. Recent efforts have focused on GPU-accelerated solvers \cite{daviet2023interactive, edwin}, complex highly curly geometry \cite{WUSHI2023, HWU2024}, real-time hair interpolation schemes \cite{hsu2024real}, and stable integration under high stiffness parameters \cite{Hsu2025}. To achieve interactive frame rates, data-driven and reduced-order models interpolate dense motion from sparse guide strands \cite{chai2014reduced, chai2017adapt} or learn runtime approximations from precomputed simulations \cite{guan2012multi}. However, running full physics models at runtime remains computationally prohibitive for real-time applications on constrained hardware. Our method instead uses a physical simulator during training, avoiding pre-generated datasets and runtime physics solves while enabling real-time strand-level predictions without data curation.

\subsection{Self-supervised Learning}
Self-supervised learning has recently emerged as an effective paradigm for modeling elastic object deformation without requiring ground-truth simulation datasets that take substantial time to generate. By minimizing physical energy potentials directly during training, self-supervised networks generalize effectively across diverse material properties, shapes, and boundary conditions. This approach has proven effective for character-driven garment simulation in both quasi-static configurations \cite{de2023drapenet} and dynamic motion sequences \cite{santesteban2022snug, grigorev2023hood, bertiche2022neural}. HyperBones \cite{srivastava2026hyperbones} distilled a larger model (HOOD \cite{grigorev2023hood}) into more compact one using a student-teacher architecture. In a similar fashion we treat the simulator as a teacher model to supervise our neural hair integrator.

For hair modeling, self-supervised methods have been applied to capture quasi-static deformations across varying poses and hairstyles \cite{stuyck2024quaffure, lin2026neuralocks}. In broader physics-informed learning, self-supervised objectives have been leveraged to model reduced-order kinematics \cite{sharp2023data, lyu2024accelerate}, enrich nonlinear modes in subspace simulation \cite{wang2024neural}, and integrate dynamics in learned subspaces \cite{li2025sca,halimi2023physgraph}. Rather than minimizing rod energies, we supervise the network directly with simulated positions, which already encode the balance between inertial and elastic response.

\subsection{Neural Hair Simulation}
Among neural hair simulators (\secref{sec:intro})\cite{zhou2023groomgen, stuyck2024quaffure, lin2026neuralocks}, quasi-static and dynamic deformation are handled by separate models rather than a single formulation. Capture-based methods have been proposed to learn the hair motion via multiview images or videos ~\cite{wang2023neuwigs, Wang_2022_CVPR, sklyarova2026dynhair}. However, these methods still struggle to produce rich, physically plausible hair motion. HairWeaver~\cite{chang2026hairweaver} introduces a diffusion-based pipeline that animates a single human image with realistic, expressive hair dynamics. However, generation is slow, and the method is conditioned on CG hair and body --- which themselves may require physics simulation.

\subsection{Neural Time Integrator} \label{sec:neural_time_integrator}
In computational fluids, machine learning has been used to improve simulation quality for both prediction and correction tasks in Eulerian settings, including advection-diffusion systems, the Navier-Stokes equations, and Poisson problems. Unrolling trains the network on sequences by feeding its own predictions forward for several steps, so that the loss is evaluated on the states the network will actually encounter at inference rather than on reference states alone. Unrolling is commonly used to train on time sequences, as it yields stable results over long rollout horizons~\cite{list2024differentiability, List_Chen_Thuerey_2022, PhysRevFluids.6.050502, kochkov2021machine, um2020solver}. The underlying problem that a model trained on reference inputs encounters its own predictions at inference, causing errors to accumulate along the generated sequence, was identified in recurrent sequence prediction, where a curriculum that gradually replaces reference inputs with generated ones was proposed as a remedy~\cite{bengio2015scheduled}. In neural PDE solvers, the pushforward trick~\cite{brandstetter2022message} addresses the same distribution shift with less computation: the model is unrolled for a few steps without gradients, and only the final step is supervised.

The ability to interact with the physics solver during training proves to be crucial for learning highly nonlinear PDEs, significantly improving generalization capabilities. In these Eulerian formulations the network and a numerical solver alternate within the forward computation, so the solver's differentiability determines whether temporal gradients can propagate between frames. Among the variants of the unrolling techniques, unrolling with full temporal gradient is the slowest, as the backpropagation graph grows with the number of unrolling steps, even though it provides the best accuracy. In unrolling without temporal gradient, the backpropagation pass can be parallelized through the unrolling steps, enabling faster training with slightly degraded quality~\cite{list2024differentiability}. List et al.~\cite{list2024differentiability} report that variants without temporal gradients improve with horizon only up to a turning point, beyond which the accumulated cut gradients degrade the learning signal, whereas full temporal gradients tolerate substantially longer horizons. More broadly, prior work on Eulerian neural simulators reports diminishing accuracy gains beyond a system-specific characteristic timescale, together with gradient-related instabilities at long horizons~\cite{List_Chen_Thuerey_2022,um2020solver}. 

Interaction with solvers or rollouts during training has also been explored in Lagrangian settings: garment simulators are trained autoregressively over multi-step rollouts\cite{grigorev2023hood}. Particle-based simulators trained on precomputed trajectories with one-step supervision counteract rollout instability by perturbing training inputs with noise rather than by unrolling~\cite{sanchezgonzalez2020learning}. Such perturbations add negligible training time and improve short-horizon robustness, but a network's own residual is neither zero-mean nor isotropic (it is correlated along each strand and compounds across frames), so injected noise does not reproduce the input distribution encountered at inference. The pushforward trick~\cite{brandstetter2022message} avoids this by feeding the model its own predictions, but without long-range gradients. To reduce the computation required for unrolling with full temporal gradients, we adopt a third variant that applies randomized unrolling horizons, and compare all variants for strand-based hair simulation.

\section{Method} \label{sec:method}

\begin{figure*}
    \centering
    \includegraphics[width=1.0\linewidth]{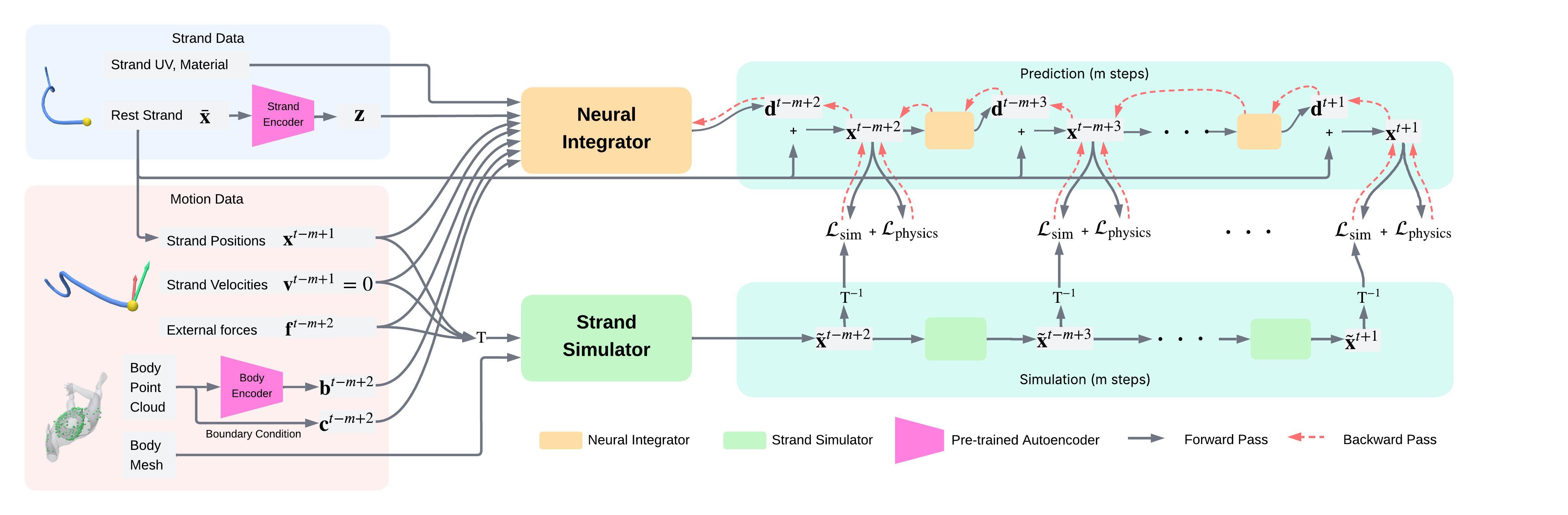}
    \caption{Our training pipeline consists of strand data and motion data for the neural integrator’s input signals. A pre-trained strand autoencoder compresses the rest-state strand data $\bar{\mathbf{x}}$ into strand latent code $\mathbf{z}$, which is then used, along with strand UV coordinates and stiffness, to condition the neural integrator. Given time $t$ and rollout horizon $m$, motion data includes strand positions $\mathbf{x}^{\tau}$, initialized by $\bar{\mathbf{x}}$, and strand velocities $\mathbf{v}^{\tau}$, which are initialized to $\mathbf{0}$. $\mathbf{x}^{\tau}$ and $\mathbf{v}^{\tau}$ are then used as the starting state for both the neural integrator and the strand simulator (initialized by rigidly transforming $\bar{\mathbf{x}}$ onto the body mesh at frame $\tau=t-m+1$). In addition, a pre-trained body autoencoder compresses the body data into body latent code $\mathbf{b}^{\tau+1}$, which provides the hair-body collision signal for the neural integrator. Together with boundary conditions $\mathbf{c}^{\tau+1}$ and external forces $\mathbf{f}^{\tau+1}$, computed in each strand root's local coordinate frame, these signals form the per-step input to the neural integrator. The neural integrator learns to predict a strand’s deformation $\mathbf{d}$ by performing $m$ prediction steps and evaluating the physics loss $\mathcal{L}_{\textup{physics}}$ as well as the $L_2$ loss $\mathcal{L}_{\textup{sim}}$ compared to the simulated strand $\tilde{\mathbf{x}}$, followed by backpropagation to update the network weights.}
    \label{fig:pipeline}
\end{figure*}

We adopt a strand-based neural network for our neural strand integrator, which keeps it lightweight and performant. Once trained, the network predicts the 3D strand state at the next frame, which is then used to autoregressively predict subsequent frames. We first introduce the dataflow of our method in \secref{sec:pipeline}. We represent all network signals in each strand’s local coordinate frame, allowing the network to learn more efficiently (\secref{sec:strand_space}). To produce a stable long-horizon rollout at inference, we adopt a simulator-in-the-loop method during training and evaluate physics losses during unrolling (\secref{sec:unrolling}). We then introduce randomized unrolling horizons, which enable both quasi-static and dynamic simulation in one network (\secref{sec:unrolling_horizons}). To learn hair-body collisions using collision loss (\secref{sec:physics_based_loss}), we condition our network on a representation-agnostic body field, generalizing to various body geometries from different 3D applications (\secref{sec:body_field}).

\subsection{Overview} \label{sec:pipeline}
In a classical physics simulation, strand states are initialized before the simulation starts, and in subsequent time steps, they are updated by time integration based on the current boundary conditions, external forces, and collision geometry. In a self-supervised machine learning workflow, however, the ground-truth simulation data is typically unavailable, or impractical to generate, unless the simulator always starts from the first frame in each training step. We therefore initialize each training rollout from a rigidly transformed rest state, followed by a short preroll with the simulator; 15 frames suffice for the strand to settle into the collision-free, draped, quasi-static configuration the network learns to reproduce (\secref{sec:unrolling_horizons}).

We first identify two types of input data for our neural integrator: strand data and motion data. Similar to Neuralocks~\cite{lin2026neuralocks}, we train an autoencoder to compress all rest-state strands $\bar{\mathbf{x}}$ into a strand latent code $\mathbf{z}\in\mathbb{R}^{32}$, which is combined with strand UV coordinates ($\mathbb{R}^{2}$) and stiffness ($\mathbb{R}^+$) to form the strand data. As in a classical physics simulator, we include in the motion data the positions $\mathbf{x}\in\mathbb{R}^{V\times3}$ and velocities $\mathbf{v}\in\mathbb{R}^{V\times3}$ for a strand with $V$ vertices, external forces $\mathbf{f}\in\mathbb{R}^{3}$, boundary conditions $\mathbf{c}\in\mathbb{R}^{3}$, and body latent code $\mathbf{b}\in\mathbb{R}^{32}$ from the collision mesh. During training, we use the unrolling strategy described above by randomly selecting a groom, a frame $t \geq m - 1$ from a body motion sequence, a body shape, a horizon $m$, and a bending stiffness. We sample the stiffness uniformly from $[20, 2000]$ and map it linearly to a multiplier $s \in [0.01, 1]$, which scales the bending potential and the simulator's bending stiffness, and is appended to the network input as a scalar. We initialize the strand state at frame $\tau = t - m + 1$ from the rest-state strand $\bar{\mathbf{x}}$, rigidly transformed onto the body mesh for our in-house simulator. At each unrolling step the simulator provides the ground-truth supervision in world space, which we map into each strand's local coordinate frame. The neural integrator instead operates entirely in this local frame, predicting the deformation $\mathbf{d}^{\tau+1}$ relative to $\bar{\mathbf{x}}$ and feeding the result into the next step untransformed. The physics loss $\mathcal{L}_{\textup{physics}}$ and the positional loss $\mathcal{L}_{\textup{sim}}$ between the simulated strand and the network prediction are accumulated (Sec.~\ref{sec:physics_based_loss}); the network then updates its weights in the backpropagation pass. Figure~\ref{fig:pipeline} depicts our training pipeline in detail.

\subsection{Strand-space Representation} \label{sec:strand_space}
In a classical time integrator, states are typically computed in world space, whereas for training a neural network, the best practice is to confine all quantities to a fixed range. To do so, we represent all data in a strand’s local coordinate frame (strand space; Figure~\ref{fig:teaser}) by applying its inverse root transformation $\mathrm{T}_{\textup{root}}^{-1}$ at frame $\tau$. Note that the boundary conditions $\mathbf{c}^{\tau+1}$ are computed from the root displacement between consecutive frames, $\mathbf{c}^{\tau+1}=\mathrm{T}_{\textup{root}}^{-1}(\mathbf{r}^{\tau+1}-\mathbf{r}^{\tau})$. The external forces $\mathbf{f}^{\tau+1}$ are likewise represented in strand space; in training they consist solely of gravity $\mathbf{g}$, $\mathbf{f}^{\tau+1}=\mathrm{T}_{\textup{root}}^{-1}\mathbf{g}$ and we find it sufficient to support arbitrary external force directions at inference time (Sec.~\ref{sec:generalizability}). We validate this choice against a world-oriented variant in Sec.~\ref{sec:ablation_studies}.

\subsection{Simulator-in-the-Loop Training} \label{sec:unrolling}
We train the integrator with multi-step unrolling, which avoids the train–test distribution shift of teacher forcing (Sec.~\ref{sec:neural_time_integrator}). Without it, compounding errors either make the rollout diverge or over-damp the hair toward a static configuration, which we measure with the rollout error and with the motion ratio and dynamic-frame count, respectively (Sec.~\ref{sec:comparison}). During training, the simulator and the neural integrator advance in parallel for $m$ frames: our in-house simulator, which solves a Cosserat rod formulation~\cite{Hsu2025} with Vertex Block Descent~\cite{chen2024vertexblockdescent}, produces the reference state at each frame, while the integrator predicts autoregressively from its own previous output. The per-frame loss is detailed in Sec.~\ref{sec:physics_based_loss}. Because reference frames are generated on the fly, no fixed dataset is precomputed per hairstyle, motion, body type, or material. In our formulation the integrator is the time-stepper: the simulator supplies positional supervision targets but never participates in the forward computation graph, so the only operator between consecutive frames is the network itself. Solver differentiability is therefore not a design consideration; what remains is how the unrolling horizon $m$ is selected and how far gradients propagate along the chain of network applications, which we address in Sec.~\ref{sec:unrolling_horizons}.

\subsection{Unrolling Horizons and Temporal Gradients} \label{sec:unrolling_horizons}
The unrolling horizon $m$ governs a trade-off between distribution-shift exposure and computational cost along two axes: how $m$ is selected at each training iteration, and how far gradients propagate along the resulting chain.

For the first axis, rather than fixing $m = M$, we sample it uniformly at each iteration, $m \sim \mathcal{U}\{1,\dots,M\}$, exposing the integrator to two regimes in a single run. At short horizons the strand state remains near the rigidly transformed rest configuration and the network learns the draped, quasi-static response to the current pose; at long horizons the state carries inertial history and the network learns to propagate dynamic motion and absorb its own compounding error. A single integrator thus acquires both modes, selected at inference by resetting or retaining the previous strand states and boundary conditions (Sec.~\ref{sec:quasistatics_and_dynamics}).

The second axis concerns gradient propagation. Unrolling $m$ frames makes the training graph a chain of $m$ applications of the integrator with weights shared along the time axis; differentiating it end to end---backpropagation through time (BPTT)---lets a later frame's loss inform an earlier prediction, but requires retaining all $m$ activations for a sequential backward pass, making the horizon the dominant driver of training memory and time. Alternatively, we apply a stop-gradient to the state passed between steps $k$, $\hat{\mathbf{x}}_{k-1} \leftarrow \operatorname{sg}[\hat{\mathbf{x}}_{k-1}]$, where $\operatorname{sg}[\cdot]$ acts as the identity in the forward pass and has zero derivative, restricting each frame's loss to that frame's network application. The forward pass is unchanged and the predicted states are numerically identical, so the distribution-shift correction of Sec.~\ref{sec:unrolling} is fully preserved; only the cross-frame gradient paths are removed. Following List et al.~\cite{list2024differentiability}, we refer to these two settings as unrolling \emph{with} and \emph{without} temporal gradients. Training without temporal gradients makes memory independent of $m$ and permits parallel per-frame backward passes. Since the two settings differ only in the availability of long-range temporal gradients, comparing them isolates whether the benefit of unrolling stems from exposure to the inference-time input distribution or from long-range gradients. 

Our default configuration is a randomized horizon with $M=10$ and temporal gradients enabled (R10). We compare it against teacher forcing, a fixed horizon, a longer horizon, and unrolling without temporal gradients in Sec.~\ref{sec:ablation_studies}.

\subsection{Physics-based Loss} \label{sec:physics_based_loss}
The objective is evaluated at every frame of the unrolled rollout rather than only at its end, and accumulated over the horizon:
\begin{equation} \mathcal{L} = \sum_{t=1}^{m} \mathcal{L}_\text{all}^{(t)}, \qquad \mathcal{L}_\text{all} = \mathcal{L}_\text{physics} + \mathcal{L}_\text{sim}, \end{equation} where $\mathcal{L}_\text{physics}$ collects the physics loss terms:
\begin{equation} \mathcal{L}_\text{physics} = \mathcal{L}_\text{collision} + \underbrace{\mathcal{L}_\text{elastic} }_{\text{optional; ablated in Sec.~\ref{sec:ablation_studies}}}, \end{equation}
where the collision barrier follows prior work~\cite{stuyck2024quaffure,lin2026neuralocks}, and the rod-specific elastic potential~\cite{lin2026neuralocks} is included for completeness but disabled in our default configuration; we find the collision term alone to be sufficient (Sec.~\ref{sec:ablation_studies}). The simulation term $\mathcal{L}_\text{sim}$ is an $L_2$ loss in strand space that supervises positions against the simulator: 
\begin{equation} \mathcal{L}_\text{sim} = w_\text{sim} \sum_\text{vertices} \left( \mathbf{x} - \mathbf{x}_\text{sim} \right)^{2}, \end{equation} where $\mathbf{x}_\text{sim} = \mathrm{T}_{\textup{root}}^{-1}\tilde{x}$ is the vertex position $\tilde{x}$ from the strand simulator transformed to strand space and $w_{\text{sim}}$ the simulation weight. This term replaces the inertia and rod-based elastic potentials of prior work (Sec.~\ref{sec:intro}): the reference trajectory already encodes the correct balance between inertial and elastic response, so the network need only reproduce it with a quadratic and well-conditioned loss. The rod-specific potential is therefore redundant rather than harmful here: the positional term already determines the elastic response, with less per-step computation. Applying $\mathcal{L}_\text{physics}$ at every unrolled frame makes it a per-step regularizer. In particular, with the collision loss $\mathcal{L}_\text{collision}$, the integrator learns to remain collision-free as a property of its own update rule rather than satisfying the constraint once at the end of a sequence, which permits long autoregressive inference with no collision projection at runtime.

We ablate the composition of $\mathcal{L}_\text{physics}$ in Sec.~\ref{sec:ablation_studies}, comparing positional supervision alone, positional supervision with the collision term, and the further addition of the rod-specific elastic potential, and report their effect on stretch violation $\mathcal{E}_\mathrm{s}$ and the intersection percentage $\mathcal{E}_\mathrm{col}$.

\subsection{Representation-agnostic Body Field} \label{sec:body_field}
As in classical time integration, our neural integrator requires a collision signal to predict collision-free strand states. Rather than feeding the collision mesh directly, we encode it as a compact body latent code, which we call the body field. The body field requires only a fixed-size set of body surface points that are temporally consistent, i.e. in correspondence across frames, so that changes in the latent code reflect body motion rather than resampling. In our implementation we obtain these points by pre-selecting 200 vertex IDs on the body mesh; for each strand we convert their 3D positions into strand space and condition an autoencoder on the resulting 200 local points, which produces the body latent code. During training, we randomly select a pose from the motion dataset along with a random body shape, and batch over strands from our groom dataset.

Because the network consumes only these local sample positions, it is unaffected by the body's mesh topology, tessellation, or the representation used to produce it. Applying the body field to a new body requires only identifying the corresponding sample locations once. Pre-selected vertex IDs are only a convenience: temporally consistent surface points can be obtained from meshes, point clouds, Gaussian splats, or volumetric representations alike, so the same trained network can be driven by any of these without retraining. Because the autoencoder always consumes a fixed number of points, its inference time is independent of the resolution of the input body geometry.

\section{Results} \label{ref:results}
We train our neural integrator on the grooms composed of the CT2Hair dataset~\cite{shen2023CT2Hair} together with synthetic hairstyles created by artists, and body motions of two input types: in-house motion capture recordings and baked animated body meshes. In all experiments, we use vertex count per strand $V$ = 25, learning rate $5 \times 10^{-4}$, simulation weight $w_{\text{sim}}=10$, and the following values for the parameters defined in Neuralocks~\cite{lin2026neuralocks}: stretching stiffness $\mathrm{k_s}$ = 100, bending-twisting stiffness $\mathrm{k_{bend\_twist}}$ = 50, collision stiffness $\mathrm{k_c}$ = 1,000, collision distance $D$ = 5~mm.

\subsection{Performance}
We build our pipeline using PyTorch and evaluate all experiments on an Intel Core i7 CPU and an NVIDIA RTX 4080 Laptop GPU. Our neural integrator is a multi-layer perceptron (MLP) consisting of 2 layers with hidden dimension 512, input dimension 217, and output dimension 72 (1.63~MB). The body autoencoder is also a 2-layer MLP, with hidden dimension 256, input dimension 600, and output dimension 32 (0.87~MB). The total inference time for predicting the deformation of a groom of 3,000 strands is 0.460~ms, comprising the body-field autoencoder (0.273~ms) and the neural integrator (0.187~ms). For predicting the same number of strands, our implementations of Quaffure and Neuralocks run at 3.550~ms and 0.203~ms, respectively. Our method is thus $7.7\times$ faster than Quaffure, while producing significantly more realistic dynamic hair motion than Neuralocks at comparable integrator runtime (Table~\ref{tab:runtime_performance_table}).

\begin{table}[]
    \centering
    \setlength{\tabcolsep}{6pt}
    \begin{tabular}{lcccc}
         & GT & Quaffure & Neuralocks & Ours \\ \toprule
        avg. time (ms)& 89.517 & 3.550 & \textbf{0.203} & 0.460\\
        $\mathcal{E}_{s} (\downarrow)$ (cm)& 0.230 & 1.527 & \textbf{0.601} & 0.850\\
        $\mathcal{E}_{col} (\downarrow)$ (\%)& 0.027 & 5.315 & 1.859 & \textbf{0.070}\\
        $\mathcal{E}_{roll} (\downarrow)$ (cm)& 0.0 & 5.686 & 3.350 & \textbf{3.339}\\
        $\mathcal{E}_{vel} (\downarrow)$& 0.0 & 0.937 & 0.956 & \textbf{0.831}\\
        avg. $\mathcal{E}_{mr} (\rightarrow1)$& 1.0 & 0.296 & 0.495 & \textbf{0.622}\\
        $\mathcal{E}_{dyn} (\uparrow)$ (frames)& 2000 & 40 & 131 & \textbf{2000}\\ \bottomrule
    \end{tabular}
    \caption{\textbf{Quantitative comparison.} Deformation metrics and runtime for predicting the next-frame state of a groom of 3,000 strands over 2,000 frames, using physics simulation (GT), prior work, and our method. We report stretch violation $\mathcal{E}_s$ ($\downarrow$), the percentage of hair-body intersections $\mathcal{E}_{col}$ ($\downarrow$), rollout error $\mathcal{E}_{roll}$ ($\downarrow$), velocity error $\mathcal{E}_{vel}$ ($\downarrow$), the motion ratio $\mathcal{E}_{mr}$ ($\rightarrow1$), and the number of consecutive dynamic frames $\mathcal{E}_{dyn}$ ($\uparrow$). Our neural integrator sustains dynamic motion over the entire sequence, and it attains the best collision, rollout, velocity, and motion-ratio scores. Best results in bold.}
    \label{tab:runtime_performance_table}
\end{table}

\subsection{Comparison} \label{sec:comparison}
For a quantitative comparison of the deformation, we measure the following metrics for both prior work and our method. All metrics except $\mathcal{E}_{mr}$ and $\mathcal{E}_{dyn}$ are computed per frame and averaged over the sequence. \begin{itemize} \item \emph{Stretch violation} $\mathcal{E}_s$: the maximum length difference over all strands. \item \emph{Collision error} $\mathcal{E}_{col}$: the percentage of hair-body intersections. \item \emph{Rollout error} $\mathcal{E}_{roll}$: the mean per-vertex $L_2$ position error against the ground truth (GT) from our in-house physics simulator. \item \emph{Velocity error} $\mathcal{E}_{vel}$: the mean per-vertex error of the body-local velocity $(\mathbf{x}^{t}-\mathbf{x}^{t-1})/\Delta t$ relative to GT. \item \emph{Motion ratio} $\mathcal{E}_{mr} = \mathrm{Motion}(N,t)/\mathrm{Motion}(\mathrm{GT},t)$,\newline where $\mathrm{Motion}(N,t)$ averages over all hair vertices the $L_2$ distance between a vertex's body-local positions at consecutive frames $t$ and $t-1$ for method $N$. Values closer to $1$ indicate motion magnitude closer to GT. Because a network can lower rollout error simply by damping motion, we read $\mathcal{E}_{roll}$ together with $\mathcal{E}_{mr}$ and treat an error reduction at low motion ratio as a failure rather than an improvement. \item \emph{Dynamic frames} $\mathcal{E}_{dyn}$: the number of consecutive frames from frame 0 before the first frame with $\mathcal{E}_{mr} < 0.35$, i.e.\ how long the hair remains ``non-rigid'' before collapsing toward a static state. \end{itemize} We report the maximum over strands rather than the mean for $\mathcal{E}_s$ because a single visibly over-stretched strand is an artifact that a per-strand average would mask. We apply the same motion sequence and body shape to two hairstyles, using our own implementations of Dynamic Quaffure and Neuralocks, for both quantitative and qualitative comparison. Table~\ref{tab:runtime_performance_table} shows the quantitative evaluation. Our method outperforms prior work on all metrics that reflect dynamic motion and reduces $\mathcal{E}_{col}$ by more than an order of magnitude relative to Neuralocks, at a moderately higher stretch violation ($\mathcal{E}_{s}$ 0.850 vs. 0.601 cm).

\subsubsection*{Dynamic Quaffure} 
\begin{figure}
    \centering
    \includegraphics[width=1.0\linewidth]{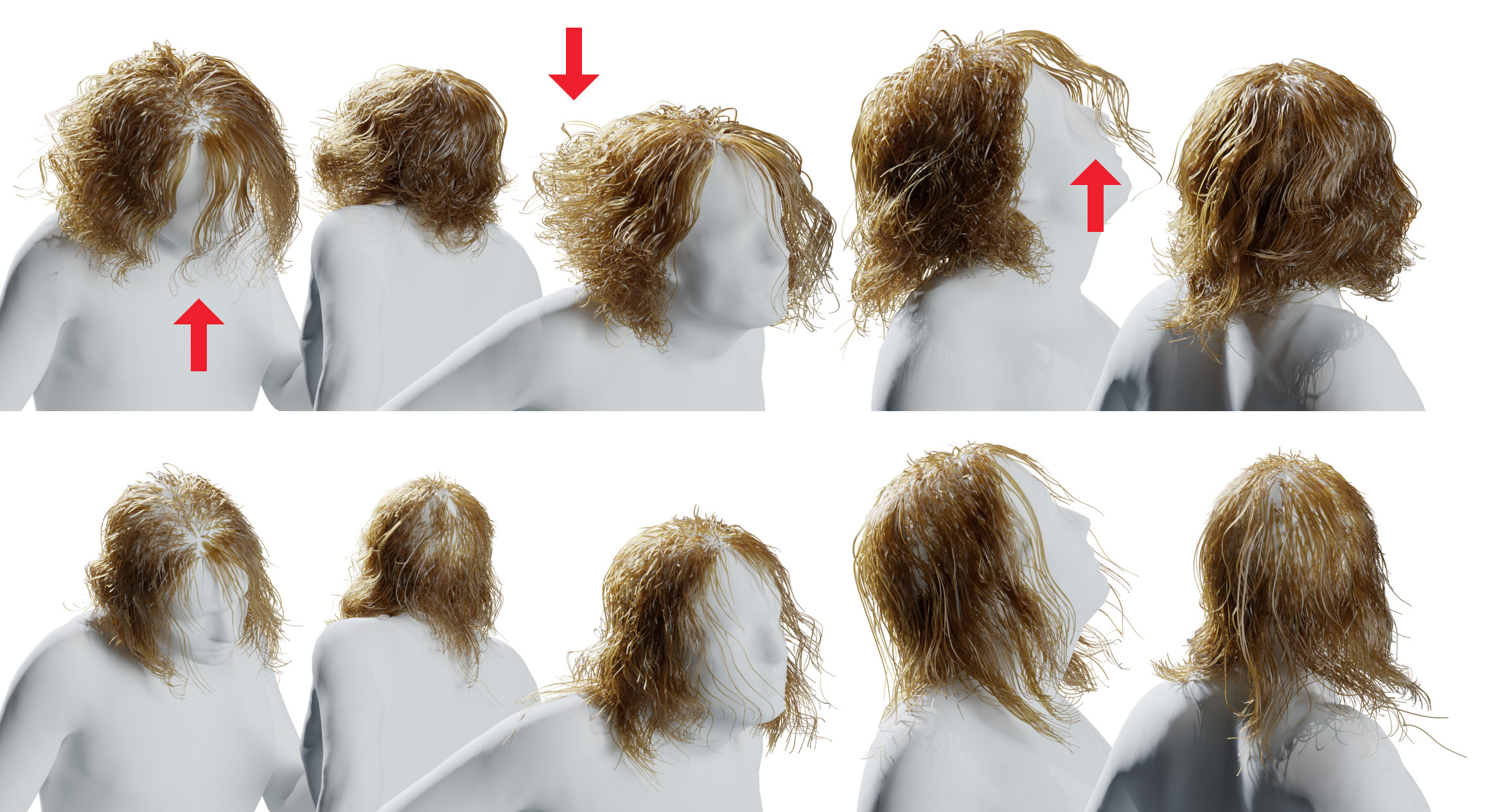}
    \caption{\textbf{Comparison with Dynamic Quaffure.} Our neural integrator produces more physically plausible hair motion (top, arrows) than Dynamic Quaffure (bottom) while running $7.7\times$ faster.}
    \label{fig:quaffure_vs_neutress}
\end{figure}

For a fair comparison, we extend Quaffure~\cite{stuyck2024quaffure} to include dynamics using the inertia potential described in Neuralocks~\cite{lin2026neuralocks}; this extension inherits the limitations of that formulation. Quaffure's single-pass, full-groom prediction caps its implementation at 4096 strands, and scaling up requires larger maps that increase both memory usage and runtime, whereas our strand-based method scales linearly with strand count. More importantly, our method produces substantially more dynamic hair motion, whereas the motion generated by Quaffure remains limited (Figure~\ref{fig:quaffure_vs_neutress}).

\subsubsection*{Neuralocks} 
\begin{figure}
    \centering
    \includegraphics[width=1.0\linewidth]{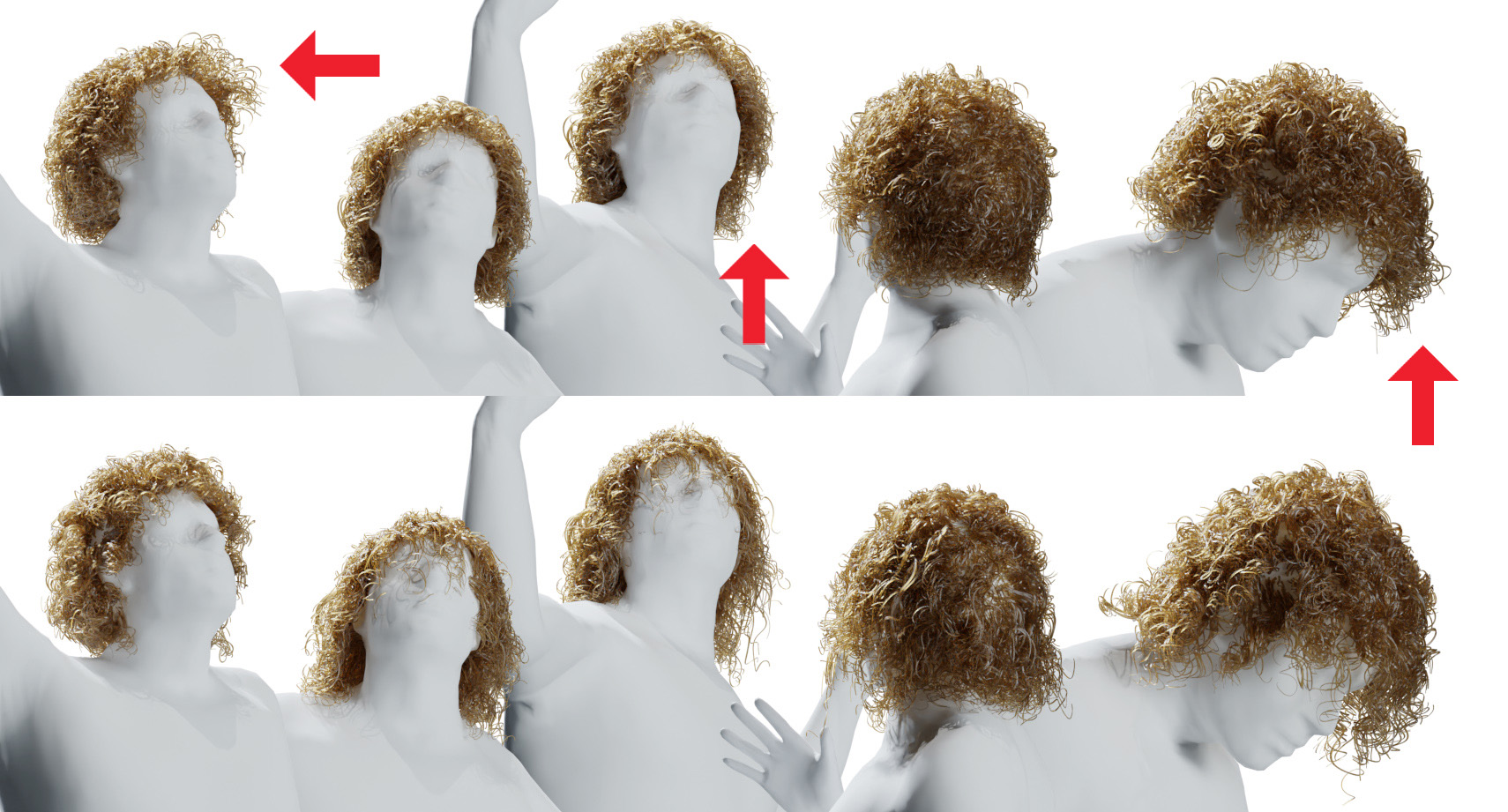}
    \caption{\textbf{Comparison with Neuralocks.} Our neural integrator produces richer individual strand motion (top, arrows) than Neuralocks (bottom), while preserving strand shape throughout the motion sequence.}
    \label{fig:neuralocks_vs_neutress}
\end{figure}

Compared with Neuralocks~\cite{lin2026neuralocks}, we observe substantially better generalization across unseen hairstyles and body motions, which we attribute to our strand-space pipeline and the simulator-in-the-loop unrolling strategy. The difference is most pronounced in sustained dynamics: Neuralocks retains less than 35\% of the reference motion after 131 frames, whereas our integrator remains dynamic for the full 2000-frame sequence (Table~\ref{tab:runtime_performance_table}, $\mathcal{E}_{dyn}$), at comparable integrator runtime (Figure~\ref{fig:neuralocks_vs_neutress}).

\subsubsection*{Physics-based Simulation} 
\begin{figure*}
    \centering
    \includegraphics[width=0.93\linewidth]{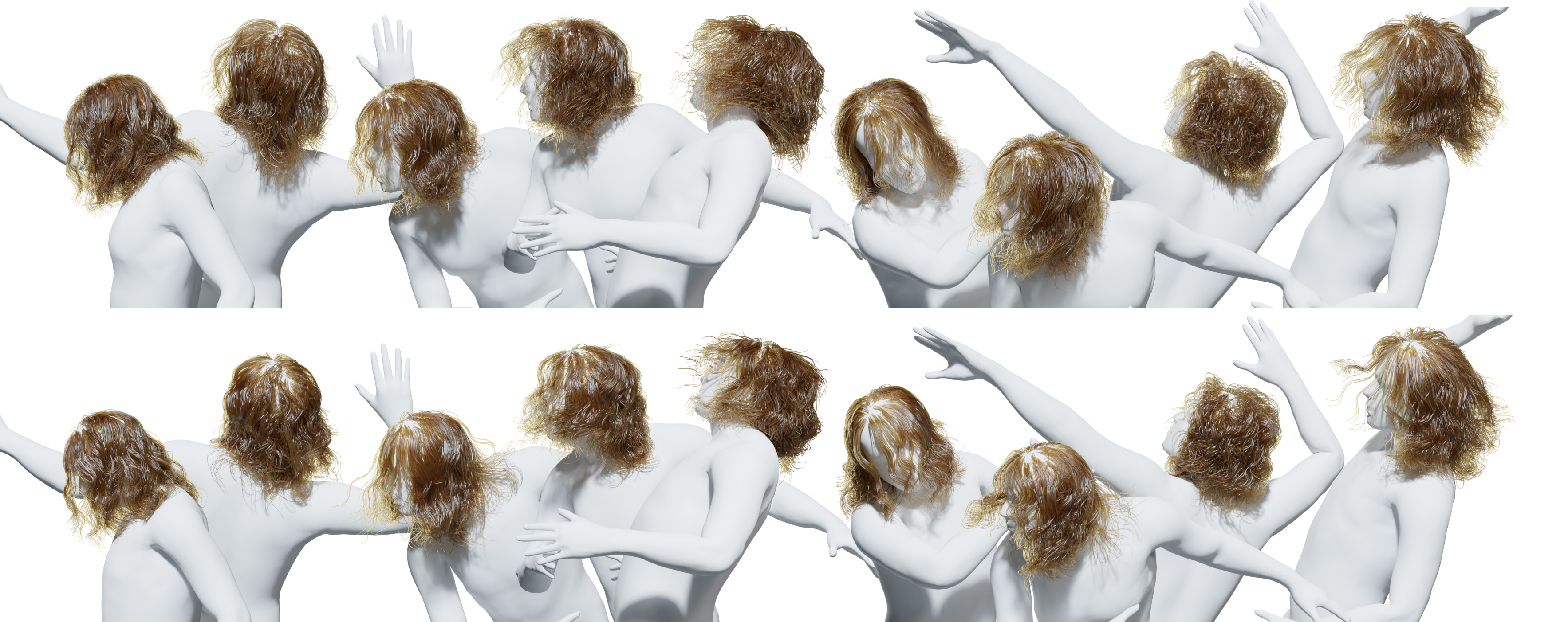}
    \caption{\textbf{Comparison with physics simulation.} Our neural integrator generates dynamic hair motion (top) comparable to that produced by our in-house strand simulator (bottom), at a fraction of the computational cost.}
    \label{fig:sim_vs_neutress}
\end{figure*}

We compare our method with classical time integration from our in-house physics simulator, which uses a Cosserat rod formulation~\cite{Hsu2025} solved with Vertex Block Descent (VBD)~\cite{chen2024vertexblockdescent} for strand dynamics and volumetric density field \cite{hadap2001modeling} for strand self-collision on the GPU. We use 20 substeps per frame and 20 VBD iterations per substep for the ground-truth simulation. Our neural integrator is more than two orders of magnitude faster than the physics simulator, which takes 89.5~ms per frame for the same 3,000-strand groom, while producing visually similar hair motion (Figure~\ref{fig:sim_vs_neutress}).

\subsection{Generalizability} \label{sec:generalizability}
To evaluate the generalizability of our neural integrator, we train the network on only a small subset of our groom and motion-capture datasets: 4 selected grooms ranging from straight to curly, including constrained hairstyles such as ponytails, together with 6 motion sequences (13,161 frames) covering indoor and outdoor activities. We then evaluate on unseen grooms, motions, and body shapes.

\subsubsection*{Stiffness} 
\begin{figure}
    \centering
    \includegraphics[width=1.0\linewidth]{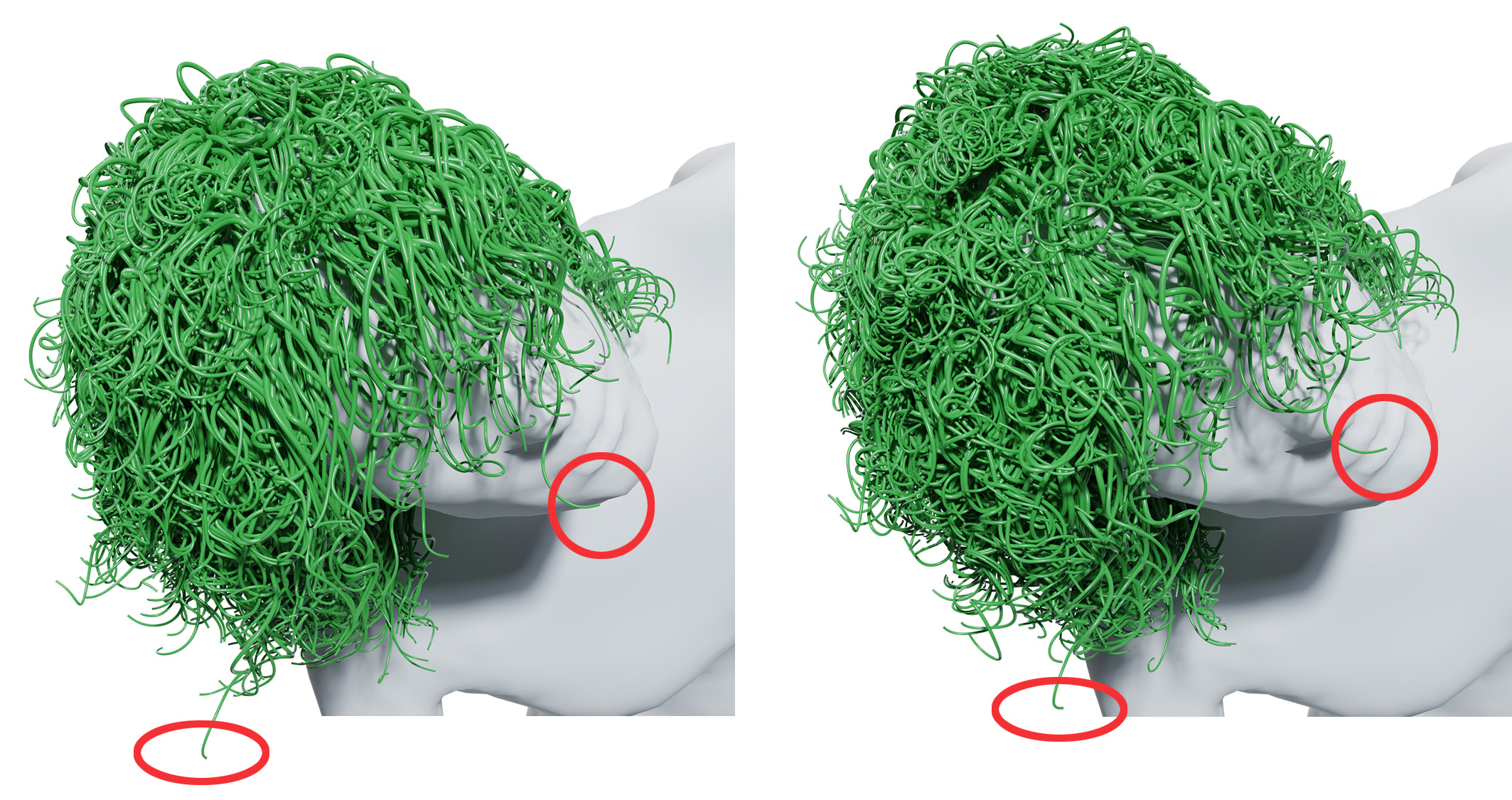}
    \caption{\textbf{Stiffness conditioning.} (Left) Lower stiffness produces sagging hair. (Right) Higher stiffness better preserves the natural curvature. Highlighted strands in circles.}
    \label{fig:stiffness}
\end{figure}

Varying stiffness multiplier $s$ (\secref{sec:pipeline}) at inference produces correspondingly different hair deformations (Figure~\ref{fig:stiffness}): low values yield sagging strands, while high values preserve the groom's natural curvature. A single trained network thus covers the full sampled range without per-material retraining.

\subsubsection*{Grooms and Body Motions} 
\begin{figure*}
    \centering
    \includegraphics[width=1.0\linewidth]{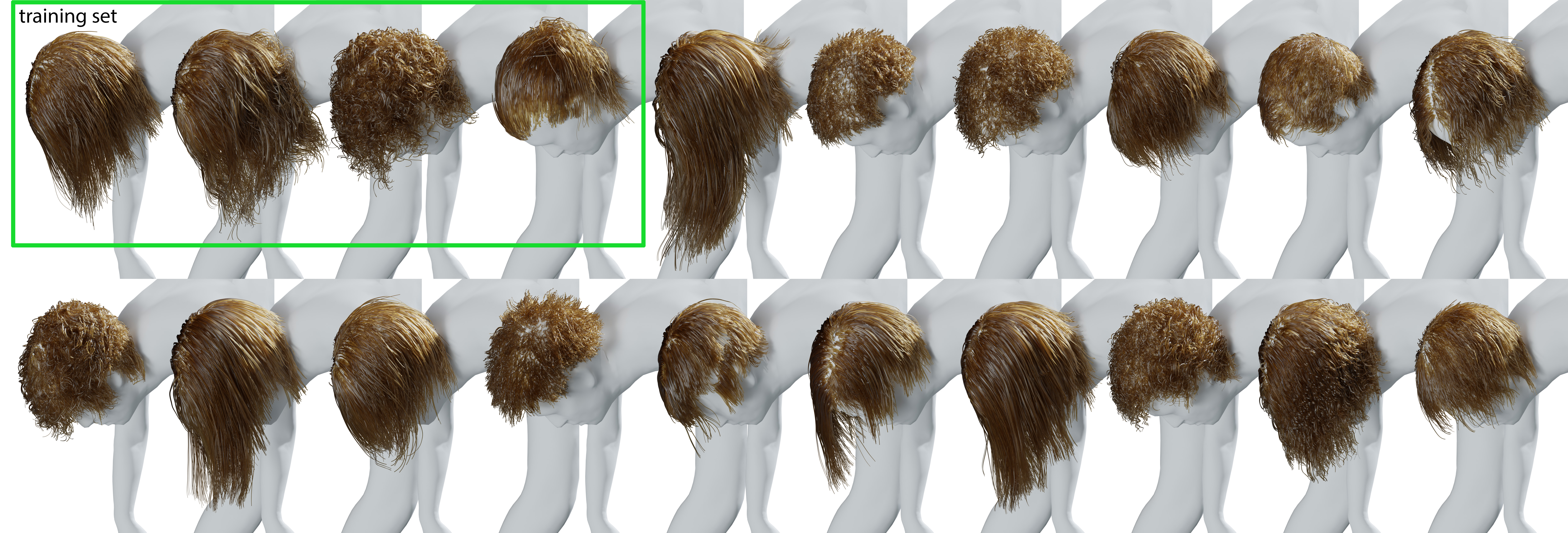}
    \caption{\textbf{Groom generalization.} Our neural integrator generates reasonable hair deformation on the grooms seen during training and generalizes well to unseen grooms.}
    \label{fig:multi_grooms}
\end{figure*}

\begin{figure*}
    \centering
    \includegraphics[width=1.0\linewidth]{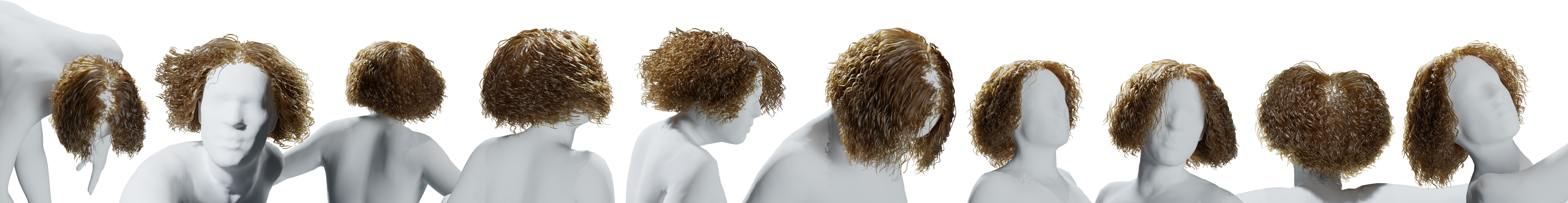}
    \caption{\textbf{Motion generalization.} Our neural integrator produces plausible hair dynamics on a wide range of body motions.}
    \label{fig:motion_generalization}
\end{figure*}

Figure~\ref{fig:multi_grooms} and Figure~\ref{fig:motion_generalization} show that our compact network generalizes to 16 unseen grooms spanning a wide range of hairstyles and dozens of unseen motion sequences, producing plausible hair deformation.

\subsubsection*{Strand Generalization} 
\begin{figure*}
    \centering
    \includegraphics[width=1.0\linewidth]{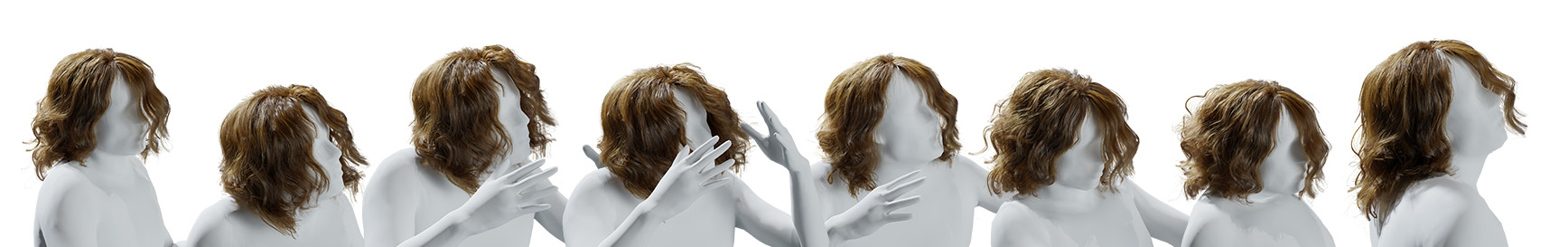}
    \caption{\textbf{Strand generalization.} Our neural integrator generalizes to full grooms without retraining.}
    \label{fig:strand_generalization}
\end{figure*}

In our experiments, we train our neural integrator on 3,000 strands per groom. Because the network operates per strand, the same trained model generalizes to full grooms of up to 120,000 strands without retraining, producing plausible deformation at 6.503~ms per frame (Figure~\ref{fig:strand_generalization}).

\subsubsection*{External Force} 
\begin{figure}
    \centering
    \includegraphics[width=1.0\linewidth]{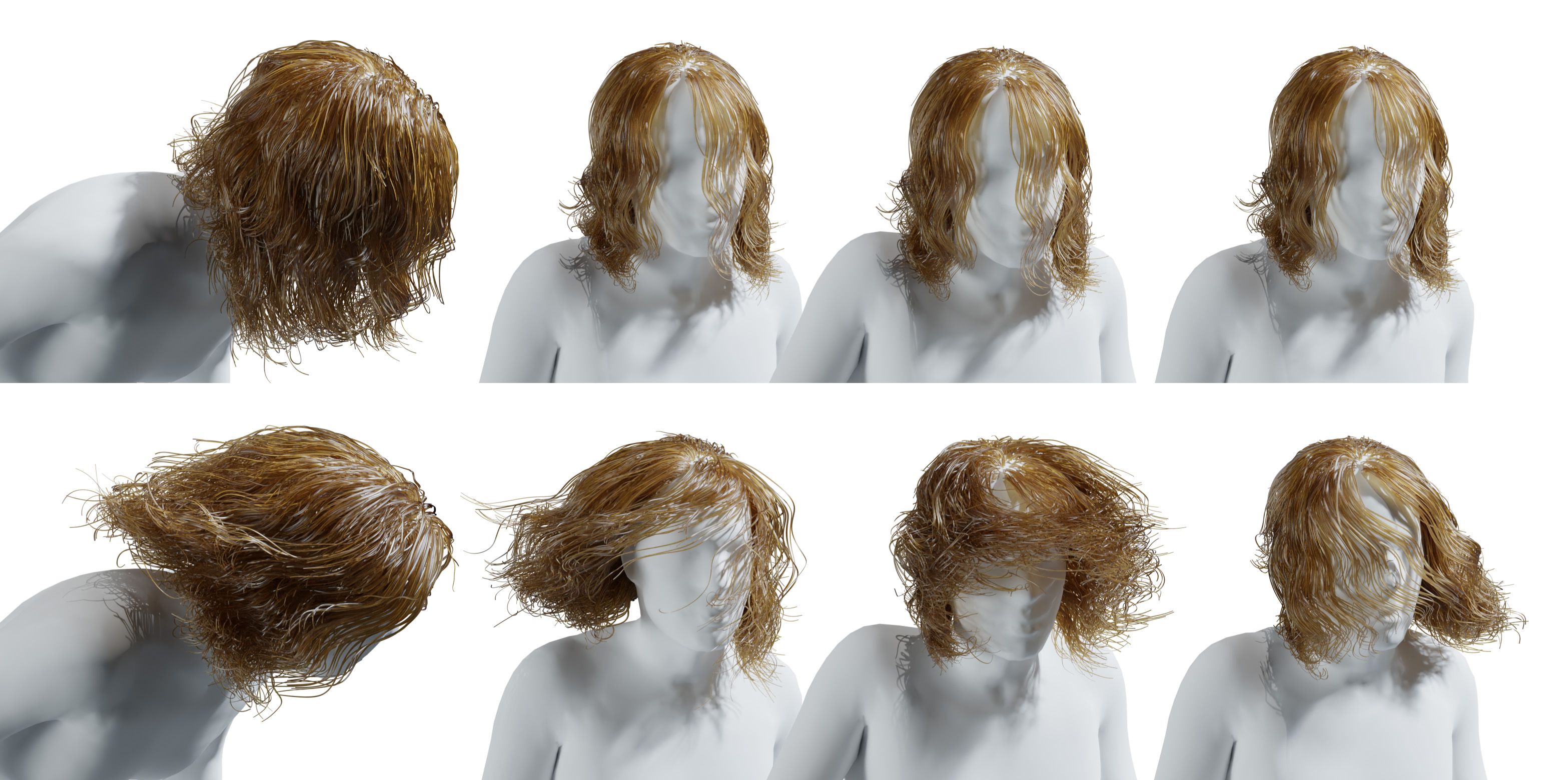}
    \caption{\textbf{External force.} Our neural integrator extends beyond gravity alone (top) to support additional external forces (bottom), such as wind.}
    \label{fig:procedural_wind}
\end{figure}

In addition to gravity, our neural integrator generalizes to arbitrary external force directions. In the example shown in Figure~\ref{fig:procedural_wind}, we add a time-varying wind force on top of gravity. Because the external force is expressed in each strand's local frame, training on gravity alone already covers a wide range of force directions. We normalize the combined force vector before feeding it to the network, and can thereby simulate hair deformation under different force directions.

\subsection{Quasi-statics and Dynamics} \label{sec:quasistatics_and_dynamics}
\begin{figure}
    \centering
    \includegraphics[width=1.0\linewidth]{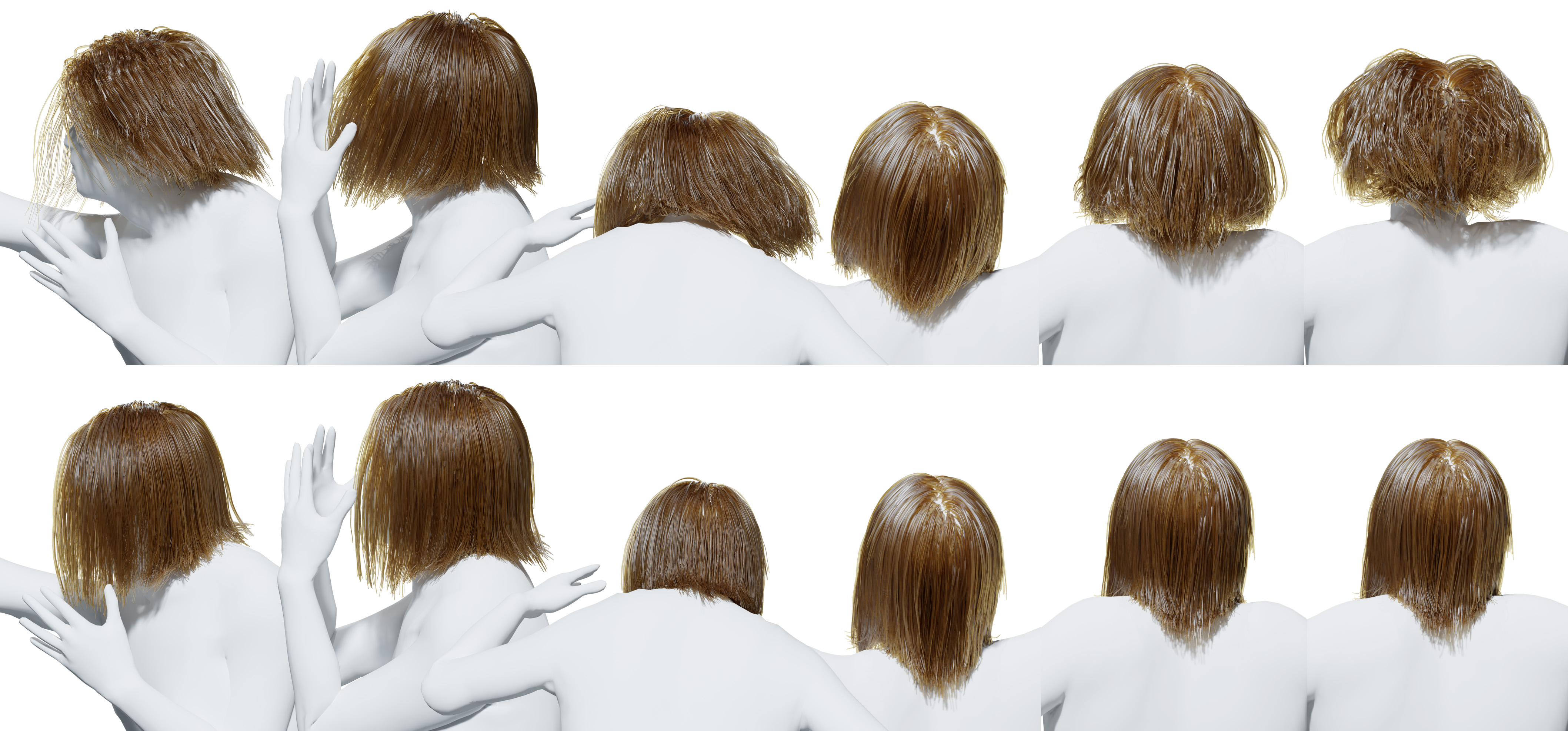}
    \caption{\textbf{Quasi-statics and dynamics.} The same network in quasi-static mode (bottom; state reset to rest each frame) and dynamic mode (top; autoregressive).}
    \label{fig:quasistatics_vs_dynamics}
\end{figure}

Our neural integrator naturally enables both quasi-static and dynamic simulation (Figure~\ref{fig:quasistatics_vs_dynamics}). During training, the shortest sampled horizon $m=1$ pairs a rest-configuration input with a reference state produced by the simulator's preroll alone; the network therefore learns the draped deformation directly. To perform quasi-static simulation at inference, the only change is to reset the neural integrator's input strand states to the rest configuration and the boundary condition to zero at each frame. When the actual strand states are used instead, the same network performs dynamic simulation.

\subsection{Ablation Studies} \label{sec:ablation_studies}
We ablate the main design choices of our method. Each ablation uses a single groom and motion sequence, with all variants differing only in the component under study.

\subsubsection*{Strand-space Representation} 
\begin{figure}
    \centering
    \includegraphics[width=1.0\linewidth]{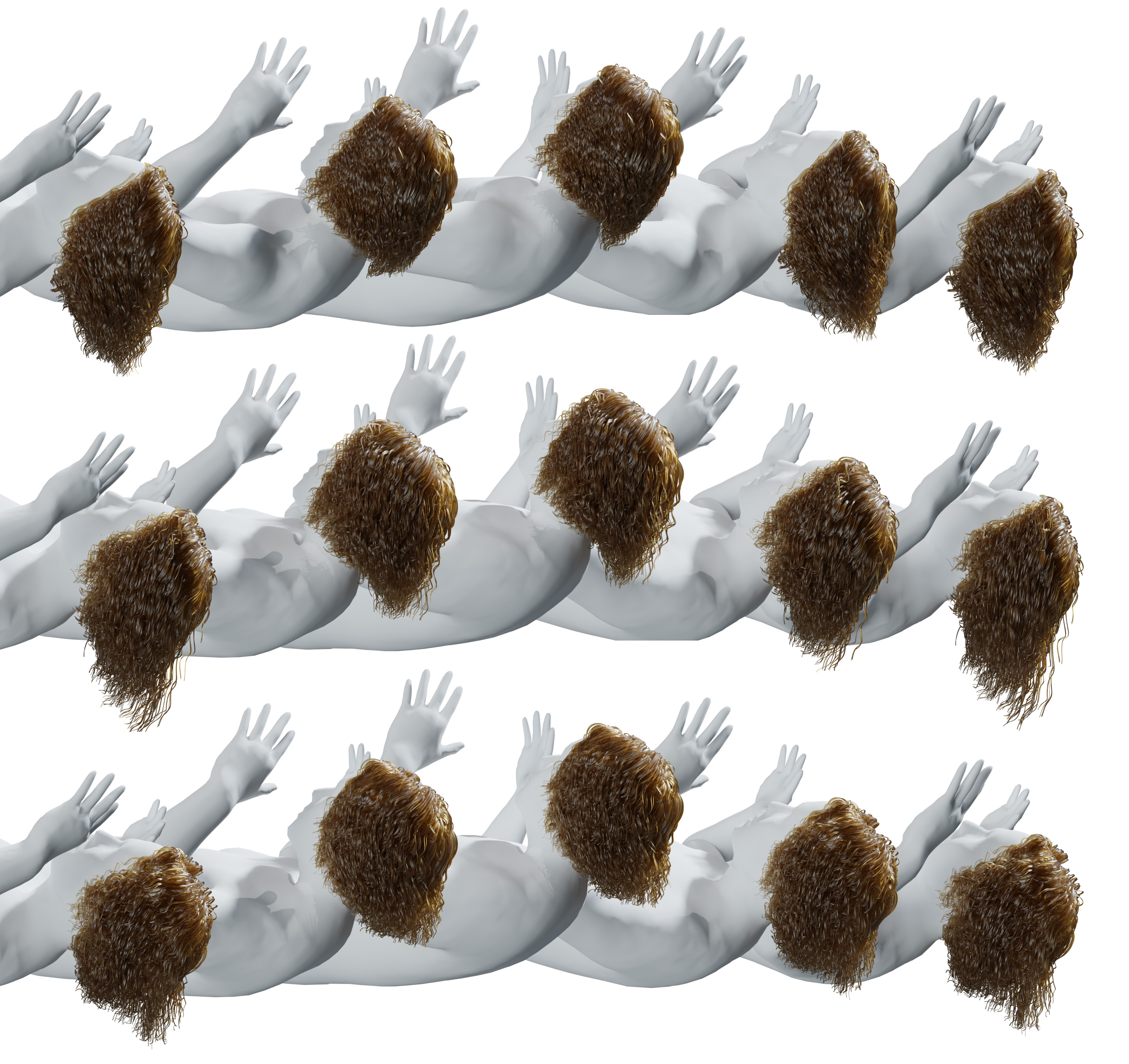}
    \caption{\textbf{Strand-space ablation.} Physics simulation (GT, top), ours in strand space (middle), and the world-space variant (bottom). The world-space variant moves nearly rigidly with the head, while ours matches the dynamic motion of the simulation.}
    \label{fig:ablation_strand_space_representation}
\end{figure}

We compare against a world-space variant that expresses all network inputs and targets relative to each strand root but in world axes; the loss, data, body field, and prediction of deformation relative to the rest strand are unchanged. The world-space variant produces nearly rigid hair with little secondary motion (avg. $\mathcal{E}_{mr}$ 0.342 vs. 0.632; $\mathcal{E}_{dyn}$ 30 vs. 800 frames), as well as substantial interpenetration ($\mathcal{E}_{col}$ 11.45\% vs. 0.001\%), whereas ours stays dynamic and nearly collision-free (Table~\ref{tab:ablation_strand_space_representation}, Figure~\ref{fig:ablation_strand_space_representation}). In world space, the strand configuration depends on head orientation, so the same physical deformation appears at many different inputs and the network settles on near-zero deformation. Strand space instead canonicalizes the strand: its rest configuration is identical regardless of orientation, and all orientation dependence is concentrated in the direction of gravity, a single vector the network can learn to respond to.

\begin{table}[]
    \centering
    \setlength{\tabcolsep}{5pt}
    \begin{tabular}{lccc}
         & GT & Strand Space (ours) & World Space\\ \toprule
        $\mathcal{E}_{s} (\downarrow)$ (cm)& 0.612 & \textbf{0.652} & 0.792\\
        $\mathcal{E}_{col} (\downarrow)$ (\%)& 0.0 & \textbf{0.001} & 11.451\\
        $\mathcal{E}_{roll} (\downarrow)$ (cm)& 0.0 & \textbf{2.054} & 2.305\\
        $\mathcal{E}_{vel} (\downarrow)$& 0.0 & 0.362 & \textbf{0.331}\\
        avg. $\mathcal{E}_{mr} (\rightarrow1)$& 1.0 & \textbf{0.632} & 0.342\\
        $\mathcal{E}_{dyn} (\uparrow)$ (frames)& 800 & \textbf{800} & 30\\ \bottomrule
    \end{tabular}
    \caption{\textbf{Strand-space ablation.} Deformation metrics for the strand-space (ours) and world-space variants on an unseen groom with 3,000 strands and an unseen motion over 800 frames, with physics simulation (GT) as reference. Both variants share the loss, data, body field, and training budget.}
    \label{tab:ablation_strand_space_representation}
\end{table}

\subsubsection*{Physics Loss} 
\begin{figure}
    \centering
    \includegraphics[width=1.0\linewidth]{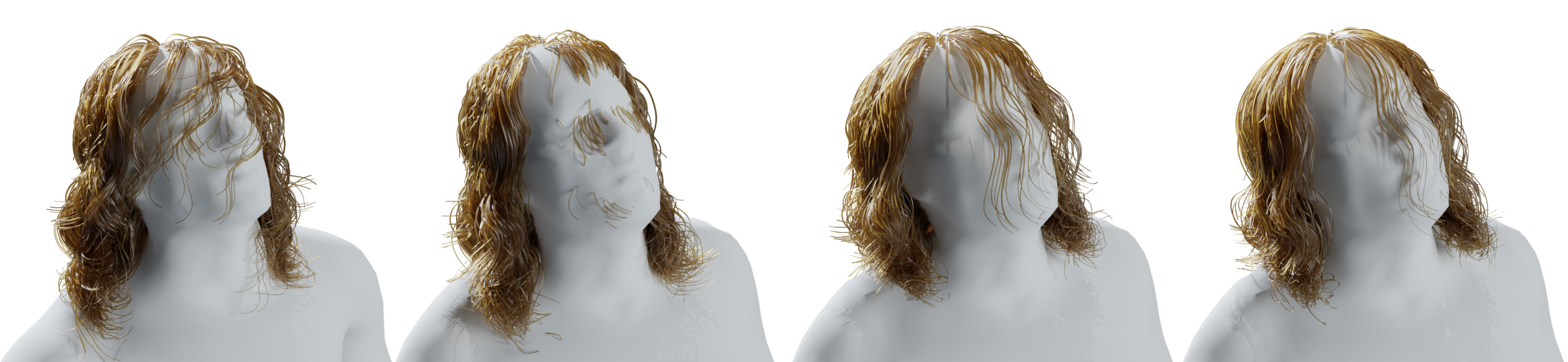}
    \caption{\textbf{Physics loss visual quality.} From left to right: physics simulation, $\mathcal{L}_\textup{sim}$, $\mathcal{L}_\textup{sim}+\mathcal{L}_\textup{collision}$, $\mathcal{L}_\textup{all}$. $\mathcal{L}_\textup{sim}+\mathcal{L}_\textup{collision}$ effectively resolves collisions while achieving comparable deformation metrics with shorter training time.}
    \label{fig:physics_loss}
\end{figure}

Table~\ref{tab:ablation_physics_loss_table} compares deformation metrics across different physics-loss configurations, and Figure~\ref{fig:physics_loss} compares the corresponding visual quality. Training with $\mathcal{L}_\text{sim}$ alone, that is without the collision loss, leaves 7.184\% of vertices inside the body. Positional supervision alone is insufficient here because the squared positional error is symmetric and penalizes a vertex displaced into the body exactly as much as one displaced away from it. Adding the collision term reduces interpenetration to 0.070\%, while reducing motion magnitude ($\mathcal{E}_\text{mr}$ 0.930 vs. 0.622). Adding the rod potentials $(\mathcal{L}_\text{all})$ recovers parts of the motion magnitude lost due to the collision barrier ($\mathcal{E}_\text{mr}$ 0.735 vs. 0.622) but doubles the interpenetration percentage and increases rollout error. We therefore adopt $\mathcal{L}_\text{sim} + \mathcal{L}_\text{collision}$, trading a small amount of motion magnitude for collision-free rollouts.

\begin{table}[]
    \centering
    \setlength{\tabcolsep}{4pt}
    \begin{tabular}{l|c|c|c|c}
         & GT & $\mathcal{L}_\text{sim}$ & $\mathcal{L}_\text{sim} + \mathcal{L}_\text{collision}$ (ours) & $\mathcal{L}_\text{all}$ \\ \toprule
        $\mathcal{T}_{train} (\downarrow)$ (s)& N/A & \textbf{1.050} & 1.320 & 1.470\\
        $\mathcal{E}_s (\downarrow)$ (cm)& 0.230 & 1.292 & \textbf{0.850} & 0.855\\
        $\mathcal{E}_{col} (\downarrow)$ (\%)& 0.027 & 7.184 & \textbf{0.070} & 0.143\\
        $\mathcal{E}_{roll} (\downarrow)$ (cm)& 0.0 & \textbf{3.157} & 3.339 & 3.648\\
        $\mathcal{E}_{vel} (\downarrow)$& 0.0 & 0.931 & \textbf{0.831} & 0.913\\
        avg. $\mathcal{E}_{mr} (\rightarrow1)$& 1.0 & \textbf{0.930} & 0.622 & 0.735\\ \bottomrule
    \end{tabular}
    \caption{\textbf{Physics loss comparison.} Deformation metrics for predicting the next-frame state of a groom of 3,000 strands over 2,000 frames, with physics simulation (GT) as the reference and networks trained with $\mathcal{L}_\textup{sim}$, $\mathcal{L}_\textup{sim} + \mathcal{L}_\textup{collision}$, and $\mathcal{L}_\textup{all}$, which additionally includes the rod elastic potentials. $\mathcal{L}_\textup{sim} + \mathcal{L}_\textup{collision}$ (ours) resolves collisions with a slight reduction in motion magnitude. Training time per step $\mathcal{T}_{train}$ is also reported. Best results in bold; for $\mathcal{E}_\mathrm{mr}$, values closest to 1 are best.}
    \label{tab:ablation_physics_loss_table}
\end{table}

\subsubsection*{Transient Time Horizon} 
\begin{figure}
    \centering
    \includegraphics[width=1.0\linewidth]{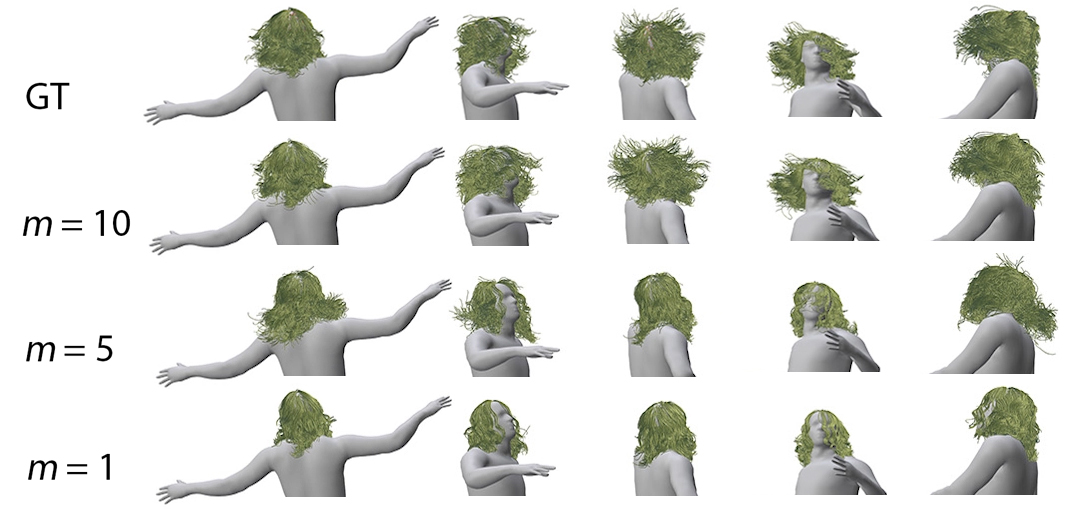}
    \caption{\textbf{Transient time horizons.} The strand states at each frame $t$ are generated by first rigidly transforming the rest-state strand onto the body mesh at frame $t-m+1$, then running the simulator for $m$ frames using our in-house strand simulator. We find that for $m \geq 10$, the resulting state closely resembles the ground truth (GT, the full simulation run from the first frame) which motivates the unrolling strategy we use during training.}
    \label{fig:horizon_experiments}
\end{figure}

We first determine how many frames of simulation are required for the strand state to become independent of its initialization. For a given horizon $m$, we generate the state at frame $t$ by rigidly transforming the rest-state strand onto the body mesh at frame $\tau = t-m+1$ and then running the simulator for $m$ frames. Figure~\ref{fig:horizon_experiments} compares the result against the ground truth, obtained by simulating from the first frame. The deformation matches the ground truth for $m \geq 10$, and we therefore set $M = 10$ for the unrolling schedules evaluated below.

\subsubsection*{Unrolling Variants} We ablate the unrolling strategy on a single groom of 3,000 strands over an 850-frame boxing sequence, denoting teacher forcing by TF, a fixed horizon $m = M$ by F$M$, a randomized horizon $m \sim \mathcal{U}\{1,\dots,M\}$ by R$M$, and the randomized schedule without temporal gradients by RNT$M$ (Table~\ref{tab:ablation_unrolling_table}). Following~\cite{sanchezgonzalez2020learning}, our TF baseline perturbs the simulator-provided input state with noise ($\sigma$ = 0.05~cm, applied to all vertices except the root) to improve robustness to its own prediction errors. Teacher forcing fails outright, confirming the distribution-shift argument of Sec.~\ref{sec:unrolling}: TF diverges after two frames, with a stretch violation an order of magnitude above any unrolled variant, so we report it as a reference point only. Among the unrolled variants, R10 reports the lowest stretch violation, rollout error, and velocity error, with less than half F10's training time per step, since a uniform schedule halves the mean horizon. F10 retains a small advantage in intersection percentage and motion ratio. Extending the horizon does not help: R30 is worse on every metric except motion ratio and takes $2.3\times$ longer per training step. R10 and RNT10 differ by under $2\%$ on rollout error, velocity error, and intersection percentage, while RNT10 trains fastest; the benefit of unrolling therefore comes from exposure to the inference-time input distribution rather than from long-range temporal gradients, consistent with the motivation of the pushforward trick~\cite{brandstetter2022message}. Temporal gradients still give a consistent refinement, most visibly in stretch violation (0.751 vs. 0.805 cm) and motion ratio (0.561 vs. 0.531), so we retain them despite roughly 10\% longer training time per step; disabling them is preferable when memory is the limiting factor. F10 is competitive on the dynamic metrics, but supervises the draped configuration in only one frame in ten, whereas R10 devotes entire iterations to it whenever $m=1$, enabling the quasi-static mode of Sec.~\ref{sec:quasistatics_and_dynamics}.

\begin{table}[]
    \centering
    \setlength{\tabcolsep}{2pt}
    \begin{tabular}{lcccccc}
         & GT & TF & R10 (ours) & R30 & RNT10 & F10\\ \toprule
        $\mathcal{T}_{train} (\downarrow)$ (s)& N/A & 0.380 & 1.152 & 2.640 & \textbf{1.048} & 2.513\\
        $\mathcal{E}_{s} (\downarrow)$ (cm)& 0.237 & 24.0 & \textbf{0.751} & 0.961 & 0.805 & 0.767\\
        $\mathcal{E}_{col} (\downarrow)$ (\%)& 0.043 & 0.833 & 0.221 & 0.321 & 0.224 & \textbf{0.161}\\
        $\mathcal{E}_{roll} (\downarrow)$ (cm)& 0.0 & 10.10 & \textbf{2.846} & 3.240 & 2.900 & 2.960\\
        $\mathcal{E}_{vel} (\downarrow)$& 0.0 & 1.515 & \textbf{0.879} & 0.947 & 0.887 & 0.899\\
        avg. $\mathcal{E}_{mr} (\rightarrow1)$& 1.0 & 1.239 & 0.561 & \textbf{0.582} & 0.531 & 0.579\\
        $\mathcal{E}_{dyn} (\uparrow)$ (frames)& 850 & 2 & 850 & 850 & 850 & 850\\ \bottomrule
    \end{tabular}
    \caption{\textbf{Unrolling strategies.} Deformation metrics for predicting the next-frame state of a groom of 3,000 strands over 850 frames, using physics simulation (GT), teacher forcing (TF, $\sigma$=0.05~cm), random horizon M=10 (R10), random horizon M=30 (R30), random horizon M=10 without temporal gradients (RNT10), and fixed horizon m=10 (F10). Our random-horizon variant (R10) performs better than other variants on most deformation metrics. Training time per step $\mathcal{T}_{train}$ is also reported. Best results in bold (excluding exploded case in TF with $\mathcal{E}_{mr}>1$).}
    \label{tab:ablation_unrolling_table}
\end{table}

\section{Limitations and Future Work} \label{sec:limitations_future_work}
Our method generalizes well to short and mid-length hair. Long hairstyles are more difficult for two reasons. First, every training window begins from a draped configuration rather than from an arbitrary point in an ongoing motion, so the network is never exposed to the large, low-frequency swings characteristic of long hair. We plan to investigate alternative initialization strategies that cover a wider range of trajectories. Second, we restrict the body field to the head and shoulder region for performance reasons, so strands extending below that region receive no collision signal. Supporting long hairstyles would require extending the body-field coverage.

Our integrator produces more dynamic motion than prior neural methods, but it consistently under-predicts motion magnitude, reaching a motion ratio of 0.622 against the reference simulation (Table~\ref{tab:ablation_physics_loss_table}). As discussed in Sec.~\ref{sec:ablation_studies}, this stems primarily from the collision barrier rather than from the positional loss. Future work could sweep the barrier stiffness and margin to characterize the trade-off between motion magnitude and interpenetration, or adopt a collision term without this directional bias.

Self-collisions can be modeled as in prior work, but they couple strands and would forfeit the per-strand independence that underlies our linear scaling and density independence. We also observe limited improvement in deformation quality in our experiments  considering self-collision. We therefore omit it in this work. This also excludes friction, which arises at these contacts and plays an important role in hair dynamics, particularly in longer hairstyles. In future work we plan to investigate friction conditioning and its effect on the temporal evolution of hair motion.

\section{Conclusion} \label{sec:conclusion}

We presented a neural time integrator for strand-based hair simulation that combines simulator-in-the-loop supervision with randomized-horizon unrolling. Rather than relying on precomputed simulation datasets or directly optimizing highly nonlinear rod-based elastic and inertial objectives, our method learns next-step hair states using positional supervision generated by a simulator during training. This substantially simplifies the learning problem while retaining the dynamic behavior of a classical hair simulator.

A key component of our approach is the representation of network inputs in strand-local coordinates, which reduces the complexity and variability of the input space. Together with simulator-in-the-loop unrolling, this enables a compact strand-based network to generalize across hairstyles, body shapes, and motions. In addition, our representation-agnostic body conditioning decouples the learned integrator from a particular body representation. Despite being trained using only a limited set of grooms and motions, the resulting model remains stable during long autoregressive rollouts and produces dynamic hair motion comparable in quality to classical time integration. Because the shortest sampled horizons supervise the draped configuration directly, the same trained network also performs quasi-static simulation selected at inference with no mode-specific training.

Overall, our results suggest that simulator-in-the-loop training provides a practical alternative to both fully physics-driven integration and conventional simulation-data-driven learning for strand-based hair. By combining the generalization and efficiency of a learned model with direct supervision from an underlying simulator, our approach provides a step toward efficient, general-purpose neural hair simulation suitable for applications with strict computational limits.

\bibliographystyle{ACM-Reference-Format}
\bibliography{ref}

\end{document}